\documentclass[%
reprint,
superscriptaddress,
 amsmath,amssymb,
prl,
longbibliography,
]{revtex4-2}

\usepackage{romannum}
\usepackage{graphicx}
\usepackage{dcolumn}
\usepackage{bm}

\usepackage{siunitx}

\usepackage{xcolor}
\definecolor{myyellow}{RGB}{255, 214, 3}
\definecolor{mypink}{RGB}{255, 151, 199}
\definecolor{mypink2}{RGB}{255,128,128}
\definecolor{myblue}{RGB}{1, 112, 189}
\definecolor{myblue2}{RGB}{85,153,255}

\usepackage{tikz}
\usetikzlibrary{shapes.geometric,shapes.symbols}
\newcommand{\filledbox}[1]{\tikz{\path[draw=#1,fill=#1] (0,0) rectangle (0.2cm,0.2cm);}}
\newcommand{\tikzcircle}[2][red,fill=red]{\tikz[baseline=-0.5ex]\draw[#1,radius=#2] (0,0) circle ;}
\newcommand{\tikzsymbol}[2][circle]{\tikz[baseline=-0.75ex]\node[inner
sep=2pt,shape=#1,draw,#2]{};}

\begin{document}

\title{\large{Thin-Film-Mediated Deformation of Droplet during Cryopreservation}}

\author{Jochem G. Meijer}
\email[]{j.g.meijer@utwente.nl}
\thanks{Contributed equally}
\affiliation{Physics of Fluids group, Max Planck Center Twente for Complex Fluid Dynamics, Department of Science and Technology, Mesa+ Institute and J. M. Burgers Center for Fluid Dynamics, University of Twente, P.O. Box 217, 7500 AE Enschede, The Netherlands}

\author{Pallav Kant}%
\email[]{kantpallav88@gmail.com}
\thanks{Contributed equally}
\affiliation{Physics of Fluids group, Max Planck Center Twente for Complex Fluid Dynamics, Department of Science and Technology, Mesa+ Institute and J. M. Burgers Center for Fluid Dynamics, University of Twente, P.O. Box 217, 7500 AE Enschede, The Netherlands}

\author{Duco van Buuren}%
\affiliation{Physics of Fluids group, Max Planck Center Twente for Complex Fluid Dynamics, Department of Science and Technology, Mesa+ Institute and J. M. Burgers Center for Fluid Dynamics, University of Twente, P.O. Box 217, 7500 AE Enschede, The Netherlands}

\author{Detlef Lohse}%
\email[]{d.lohse@utwente.nl}
\affiliation{Physics of Fluids group, Max Planck Center Twente for Complex Fluid Dynamics, Department of Science and Technology, Mesa+ Institute and J. M. Burgers Center for Fluid Dynamics, University of Twente, P.O. Box 217, 7500 AE Enschede, The Netherlands}
\affiliation{Max Planck Institute for Dynamics and Self-Organization, Am Faßberg 17, 37077 Göttingen, Germany}

\begin{abstract}
Freezing of dispersions is omnipresent in science and technology. While the passing of a freezing front over a solid particle is reasonably understood, this is not so for soft particles.
Here, using an oil-in-water emulsion as a model system, we show that when engulfed into a growing ice front, a soft particle severely deforms.
This deformation strongly depends on the engulfment velocity $V$, even forming pointy-tip shapes for low values of $V$.
We find such singular deformations are mediated by interfacial flows in \textit{nanometric} thin liquid films separating the non-solidifying dispersed droplets and the solidifying bulk.
We model the fluid flow in these intervening thin films using a lubrication approximation and then relate it to the deformation sustained by the dispersed droplet.

\end{abstract}


\maketitle

Solidification of a liquid seeded with insoluble particles is pertinent to numerous natural and industrial processes, including metal processing \cite{deville2006freezing, deville2007ice, deville2008freeze, deville2010freeze}, cryopreservation of biological cells/tissues \cite{korber1988phenomena, muldrew2004water, bronstein1981rejection}, food engineering (ice cream making) \cite{goff1997colloidal, ghosh2008factors,crilly2008designing}, and swelling of wet ground (frost-heaving) in colder regions \cite{peppin2013physics, peppin2011frost, rempel2010frost}.
The key feature of the solidification in such complex systems is the interaction between dispersed particles and a growing crystal (or polycrystalline structure; for short, we will use "crystal" in this paper).
When a growing crystal encounters a foreign particle suspended in a melt, the moving solidification front (a solid-liquid interface) either deforms around the particle to trap it within the crystal, or rejects it \citep{shangguan1992analytical, park2006encapsulation, dedovets2018five}.
The type of behaviour that prevails at a given solidification velocity therefore, has important implications for the microstructure of the newly formed solidified material.
The interaction between a dispersed particle and the moving solidification front is dictated by an interplay among van der Waals interactions, thermal conductivity differences between the particle and the melt, solid–liquid interfacial energy, and the density change caused by the liquid–solid phase transition \cite{uhlmann1964interaction, bolling1971theory, chernov1976theory, gilpin1979model, gilpin1980theoretical, rempel2001interfacial, rempel2004premelting, rempel1999interaction, israelachvili2011intermolecular}.
In case of a droplet or a bubble suspended in the melt, this local dynamics is amplified by thermo-capillary effects \cite{park2006encapsulation, young1959motion, geguzin1981crystallization}.

In this Letter, we report a surprising and highly interesting solidification behavior that emerges when a liquid seeded with droplets of another immiscible liquid (emulsion) solidifies.
We demonstrate that the dispersed non-solidifying droplet sustains varying degree of deformations, which strongly depend on the engulfment velocity $V$, remarkably even forming a pointy tear-like shape for extremely slow engulfment.
We find such deformation to be mediated by interfacial flows in the intervening thin liquid films that separate the dispersed droplet and the solidifying bulk.
These findings are extremely relevant for exerting better control over cyropreservation procedures of food emulsions and bio-specimen, where mechanical deformations of the dispersed medium dictate the freeze-thaw stability of the solidified material.
More specifically, our findings quantitatively explain why - perhaps counterintuitively - faster freezing is less intrusive for the shape of the frozen soft structures.

\begin{figure}[b] 
\includegraphics[width=0.5\textwidth]{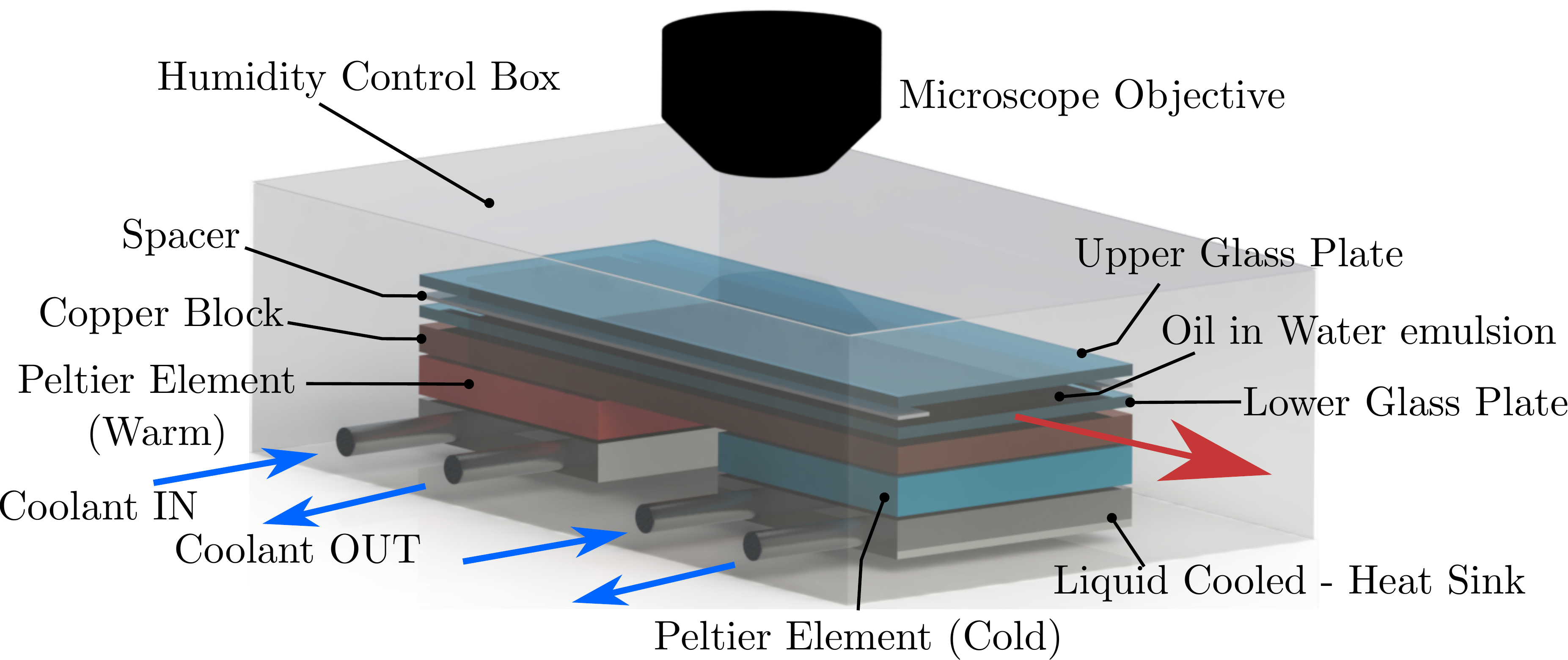}
\caption{Schematic of the experimental setup. The red arrow indicates the direction of sliding of the Hele-Shaw cell over the copper block.}
\label{fig:1}
\end{figure}
\begin{figure*} 
\includegraphics[width=0.9\textwidth]{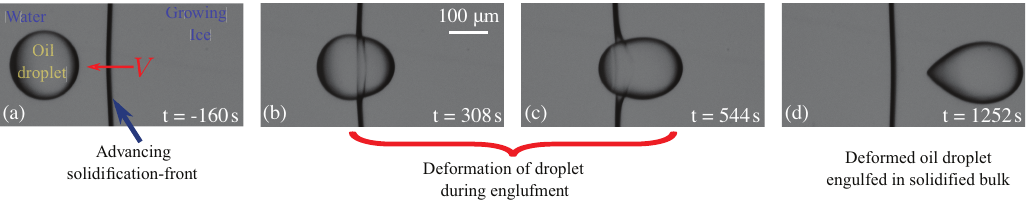}
\caption{(a)-(d) Top-view snapshots capturing the engulfment of a \SI{5}{cSt} silicone oil droplet submerged in water by an advancing solidification-front ($V =\SI{0.4}{\micro \meter \per \second}$), i.e. ice, first approaching and then passing over the droplet (Supplementary Movie 1). 
At $t=0$ the drop makes first contact with the front.
The applied temperature gradient $G$ is about $10\,\mathrm{K\,cm^{-1}}$. 
The droplet remains deformed in the solidified bulk after engulfment throughout the entire experiment (typically $\sim$ 4 hours).}
\label{fig:2}
\end{figure*}

We investigate the interaction between an advancing water-ice freezing front and silicone oil droplets in a horizontal Hele-Shaw cell of thickness 200\,$\mu$m, filled with an oil-in-water emulsion.
The emulsion is prepared using a microfluidic setup, as explained in Ref.\cite{van2021feedback}.
To ensure the stability of the emulsion, a surfactant TWEEN80 (Sigma-Aldrich, Germany) is added to the water (0.5 vol\% initially). 
Since surfactants are known to affect freezing of complex liquids \cite{dedovets2018five,kao2009particle,tyagi2022solute}, the emulsion sample is diluted after fabrication to achieve low surfactant concentration (0.01 vol\%).
As a consequence, any such solutal effects are irrelevant for our experiments.
In each experiment, the Hele-Shaw cell is translated at constant velocity $V$ through a fixed temperature gradient $G = \nabla T$ using a linear actuator (Physik Instrumente, M-230.25) such that the position of the moving solidification front remains fixed in the lab frame of reference. 
The temperature gradient $G \sim 10\,\mathrm{K\,cm^{-1}}$ is applied over the entire length of the Hele-Shaw cell by placing it over a copper block with cold and warm ends, maintained at fixed temperatures using Peltier elements.
Here we report experiments for different advancing velocities that ensure engulfment \cite{rempel1999interaction}, which were achieved by changing the thermal gradient $G$; for more details we refer to the Supplemental Material.
The engulfment process of a droplet into growing ice is recorded in top-view through a camera connected to a long working distance lens (Thorlabs, MVL12X12Z), and diffused back lighting. A schematic of the experimental setup is given in Fig.\,\ref{fig:1}.  

\begin{figure}
\includegraphics[width=0.25\textwidth]{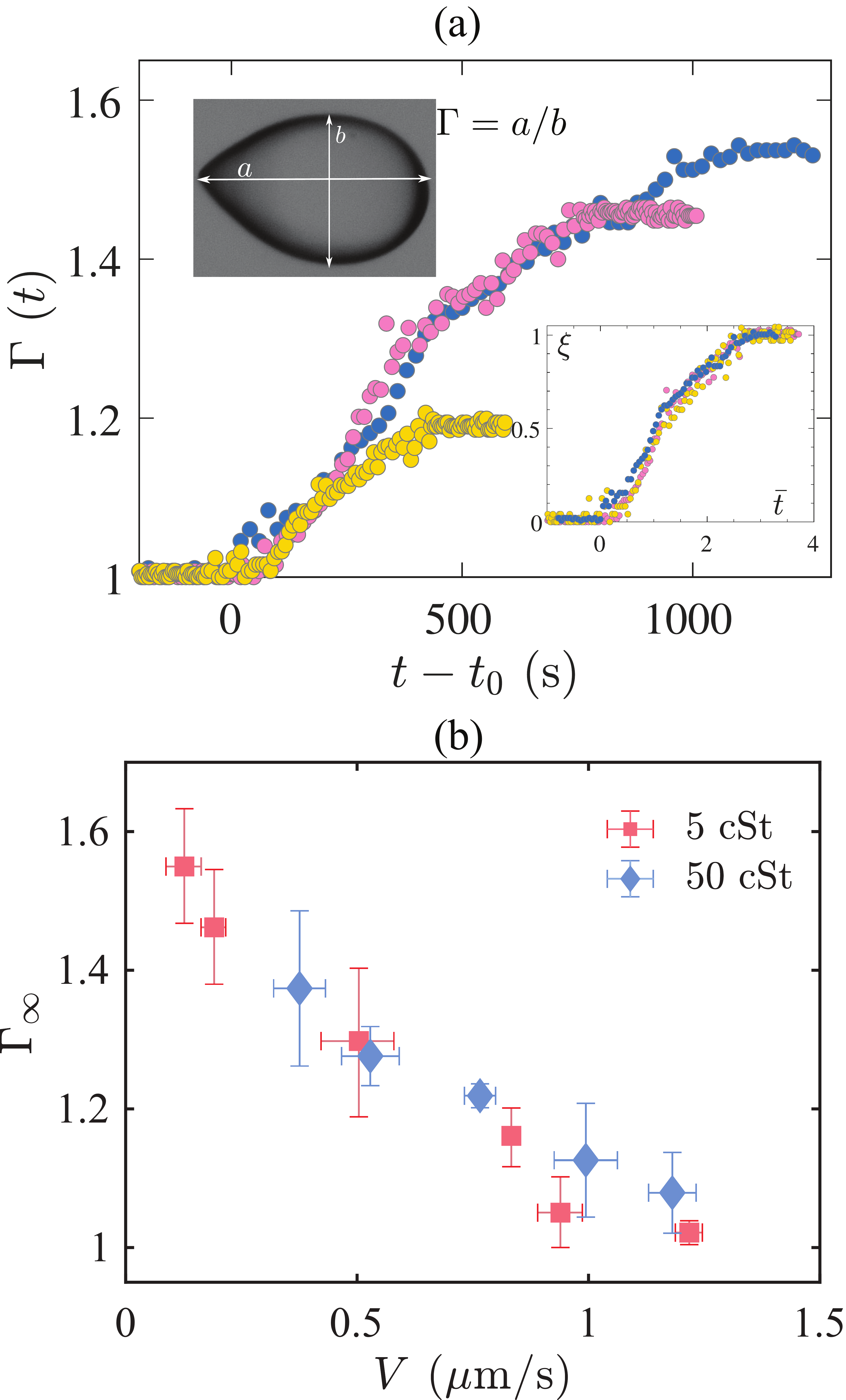}
\caption{(a) Aspect ratio $\Gamma (t)$ as a function of time $t$ for advancing velocities  $V = $ (\tikzcircle[gray, fill=myblue]{2.75pt}) $ \SI{0.2}{\micro \meter \per \second}$ (\tikzcircle[gray, fill=mypink]{2.75pt}) $ \SI{0.4}{\micro \meter \per \second}$ (\tikzcircle[gray, fill=myyellow]{2.75pt}) $ \SI{0.6}{\micro \meter \per \second}$. The solidification-front makes first contact with the droplet at $t = t_0$. 
The inset shows the universality of the engulfment process, where $\xi = (\Gamma(t) - 1)/(\Gamma_{\infty} - 1)$ and $\bar{t} = (t-t_0)V/R$. 
(b) Final aspect ratio $\Gamma_{\infty}$ of the deformed droplet in the solidified bulk as a function of the solidification front advancing velocity $V$ for two kinematic viscosities $\nu = $(\protect \filledbox{mypink2}) $\SI{5}{cSt} $ and (\tikzsymbol[diamond]{fill=myblue2}) $ \SI{50}{cSt}$. 
}
\label{fig:3}
\end{figure}

The sequence of images in Fig.\,\ref{fig:2} depicts typical deformations experienced by an oil droplet of radius $R \sim \SI{90}{\micro \meter}$ while being engulfed in the solidifying bulk at a constant advancing velocity $V$ (Supplementary Movie 1).
During the engulfment process, the initially planar solidification interface, approaching the droplet at a constant velocity $V \sim \SI{0.4}{\micro \meter \per \second}$, deforms locally to envelop the droplet.
Interestingly, this causes the droplet to squeeze laterally and stretch longitudinally at the opposite free end.
This deformation of the droplet is characterized by its instantaneous aspect ratio $\Gamma (t)$, defined as the ratio of maximum longitudinal and lateral dimensions.
Deformation curves shown in Fig.~\ref{fig:3}a highlight scenarios when a droplet is engulfed into the solidifying bulk at different engulfment velocities.
In all cases, the droplet starts to deform soon after the advancing solidification front makes contact with it at $t = t_0$, and steadily stretches to its equilibrium state characterised by aspect ratio $\Gamma_\infty$ when completely engulfed.
Remarkably, the extent of deformation sustained by the droplet shows a crucial dependence on the advancing velocity $V$ of the solidification front and hence on the applied temperature gradient $G$.
Note that a simple scaling analysis that balances the rate of heat released during solidification ($\rho \, \mathcal{L} \, V$) with its conduction through the solidified bulk ($k_s\,G$) dictates $V \propto G$; here $\rho$ and $\mathcal{L}$ are the density and the latent heat of water, $k_s$ is the thermal conductivity of ice.
At lower engulfment velocities (low $G$), the droplet severely stretches longitudinally into a tear-like shape, whereas, at increased engulfment velocities (high $G$), the droplet retains its initial spherical shape.
The decreasing trend of $\Gamma_\infty$ with advancing velocity $V$ of the solidification front is shown in Fig.\,\ref{fig:3}b for two kinematic viscosities. The latter does not seem to significantly influence the observed behaviour.
It is important to highlight that during an experiment, which typically lasts $\sim$ 4 hours, the deformed droplet retains its pointy shape once encapsulated into the solidified bulk. 

The physical mechanism that causes the deformation of a droplet during the encapsulation process is intimately connected to the phenomenon known as interfacial premelting \citep{dash1995premelting, dash2006physics, wettlaufer2006premelting}. 
It is known that at constant pressure, the melting of a bulk solid occurs at a fixed temperature $T_{\mathrm{m}}$. 
However, as the solid–liquid interface deforms around the droplet, the local melting temperature of the surrounding solidifying phase reduces below $T_{\mathrm{m}}$ by an amount proportional to the local curvature of the particle ($1/R$), and the surface energy of the solidification front $\sigma_\mathrm{sl}$. 
This reduction in local melting temperature at the expense of interfacial energy is referred to as the Gibbs-Thomson effect \cite{perez2005gibbs}.
Moreover, when the distance between the advancing solidification front and the droplet reduces to extremely low values, less than a micrometer, van der Waals interactions become significant, giving rise to a disjoining pressure $\Pi$ \cite{derjaguin1955definition}, which causes a further local reduction of the melting temperature.
Therefore, the overall reduction in the local melting temperature due to the geometrical shape of the droplet and intermolecular interactions is given by \cite{wettlaufer1996theory}:
\begin{equation}
\Delta T_\mathrm{m} = T_\mathrm{m} \left[\frac{\sigma_\mathrm{sl}}{\rho_s \mathcal{L} \, R} + \left(\frac{\lambda}{H}\right)^{\eta} \right], 
\label{eq:meltingTemp}
\end{equation} 

where $\rho_s$ is the density of the solid, $\mathcal{L}$ is the latent heat, $H$ is the distance between the droplet interface and the front, $\lambda$ the disjoining pressure length scale, and the exponent $\eta$ depends on the type of intermolecular forces that dominate; for non-retarded van der Waals
interactions $\eta = 3$ \cite{wettlaufer1996theory}.
Note that electrostatic contributions are not taken into account.
Crucially, due to the reduction in melting temperature of the bulk in the vicinity of the droplet, a thin liquid film of thickness $H(X)$ persists in the gap between the droplet and the solidifying bulk.
Direct imaging of such thin films is challenging because of their small dimensions.
However, our experimental setup allows us to observe such thin-films as the solidification front passes over the droplet, see Supplemental Material.
Here, we hypothesize that it is the spatial pressure variation within this thin film that mediates the overall shape change of the droplet during its engulfment into the solidifying bulk. 
In the following, we present a model that connects the fluid flow in the enveloping thin film around the droplet to its deformation.

\begin{figure} 
\includegraphics[width=0.45\textwidth]{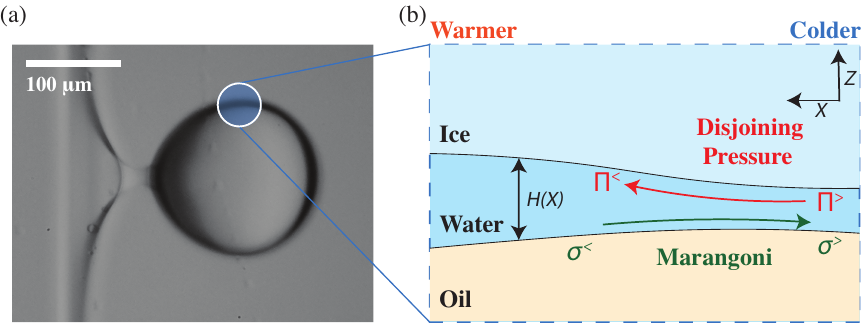}
\caption{(a) Experimental image capturing the engulfment of an oil drop (Supplementary Movie 2). (b) The schematics visualizes the competition between the disjoining pressure $\Pi$ and Marangoni effects that mediate the flow profile within the liquid film with varying $H(X)$.  
Dependent on the profile of the thin film, the pressure in the film $p_{\mathrm{film}}$ varies, which determines the local curvature of the oil-water interface and eventually the overall shape of the drop.}
\label{fig:4}
\end{figure}

We model the fluid-flow in thin intervening liquid films using lubrication theory.
The main assumptions of our analysis is that the flow during the engulfment process is predominantly in the direction of motion of the solidification front.
Accordingly, as illustrated in Fig.\,\ref{fig:4}b, we treat the intervening thin-film as a one-dimensional slab geometry, bounded between the oil-water interface and the growing ice.
Fluid motion within the film is mainly driven by a gradient in disjoining pressure due to the variation with film thickness ($\Pi \sim H^{-3}$) \cite{dai2008disjoining}, and Marangoni stresses at the droplet interface due to the externally applied thermal gradient $G$; see the schematic in Fig.\,4b.
By making use of the slender geometry of the system, and by applying no-slip and slip boundary conditions at the water-ice and oil-water interfaces as $U\vert_{(Z=H)} = 0$ and $\mu \partial_Z U\vert_{Z=0} - \mu' \partial_Z U'\vert_{Z=0} = \partial_X \sigma$, respectively, where $\partial_Z U' = -\partial_X \sigma/(\mu + \mu')$ \citep{young1959motion}, the net horizontal flux $q(X)$ is given by
\begin{equation}
\mu q(X) = \frac{1}{2}\frac{\mathrm{d}\sigma}{\mathrm{d} T}\,\left[1 - \frac{\mu'}{\mu + \mu'} \right]\,G\,H(X)^2 - \frac{1}{3}\, \partial_X p\,H(X)^3.
\end{equation}

Primed quantities refer to the oil phase and unprimed to the water phase. $\sigma$ is the interfacial tension of the oil-water interface, $\mu$ the dynamic viscosity, and $p = -\sigma \partial^2_X H - A/6\,\pi\,H^3$ is the pressure in the thin film in lubrication approximation.
Note that the disjoining pressure is also related to the applied thermal gradient as $A/6\,\pi\,H^3 = - \rho_s \mathcal{L}G\,X/T_\mathrm{m}$ \cite{wettlaufer2006premelting} with $A$ the Hamaker constant.
Consequently, the local mass conservation ($\partial_t h + \partial_x q = 0$) in non-dimensional form can be expressed as:
\begin{equation}
\partial_t h+ \partial_x\left[\frac{1}{3 Ca}h^3 \partial^3_x h + \frac{1}{3Br Pr}h^3 - \frac{Ma}{Pr}h^2 \right]= 0.
\label{eq:thinfilmeqs}
\end{equation} 

\begin{figure*} 
\includegraphics[width=0.8\textwidth]{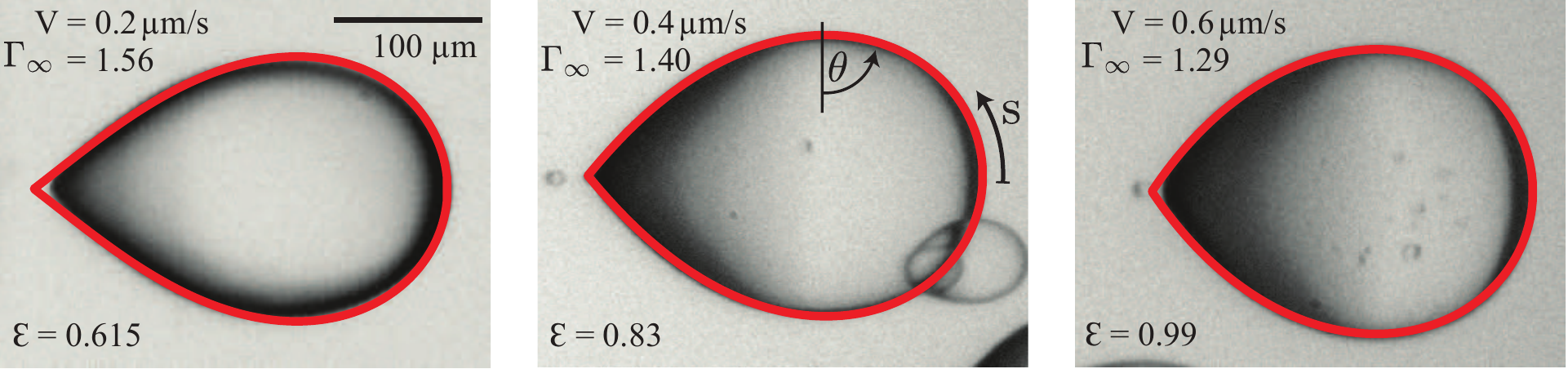}
\caption{Experimental snapshots of three deformed \SI{5}{cSt} silicone oil droplets that were engulfed at different velocities $V = \SI{0.2}{\micro \meter \per \second}, \SI{0.4}{\micro \meter \per \second}$, and $\SI{0.6}{\micro \meter \per \second}$. The applied temperature gradient $G$ is about \SI{10}{\kelvin \per \centi \meter} and the corresponding final aspect ratios $\Gamma_{\infty}$ of their final shapes are also indicated. The lines overlayed in red are the shapes generated by the pendant drop model \citep{roman2020pendulums} with corresponding values of the deformation parameter $\mathcal{E}$. }
\label{fig:5}
\end{figure*}

Note that in the above equation lowercase quantities are non-dimensionalised. The droplet radius $R$ is used to non-dimensionalize lengths, and the velocity scale is chosen as $\nu/R$.
$Ca = \mu\,\nu/\sigma R$ is the Capillary number and $Pr = \nu/\kappa$ the Prandtl number; $\nu= \mu/\rho$ and $\kappa$ are kinematic viscosity and thermal diffusivity of water, respectively.
The other two dimensionless parameters of the problem are the Brinkman number $Br$ 
and the Marangoni number $Ma$, defined as 
\begin{align}
Br = \frac{\mu \kappa T_m}{\rho_s \mathcal{L}\,G\,R^3},  \quad Ma = \frac{-(\mathrm{d}\sigma/\mathrm{d} T)\,G\,R^2}{2\mu \kappa}\left[1 - \frac{\mu'}{\mu + \mu'} \right].
\label{eq:BrandMa}
\end{align}
They indicate the strength of the fluid flow driven by intermolecular pressure and of the thermal Marangoni force, respectively.
The former expresses the (inverse) ratio of the time-scale for diffusive thermal transport ($t_{\kappa} \sim R^2/\kappa$) to the time-scale for convective thermal transport ($t_{u} \sim \mu T_{\mathrm{m}}/\rho_s \mathcal{L} G R$) that is driven by the gradient in intermolecular pressure. 
The latter provides an estimate for the (inverse) ratio of the time-scale for convective thermal transport ($t_{Ma} \sim -\mu / (\mathrm{d}\sigma/\mathrm{d}T) \, G$) driven by the thermal Marangoni force and again the time-scale for diffusive transport.
Although it is difficult to predict \textit{a priori} whether intermolecular interactions dominate over thermal Marangoni effects or vice versa, by changing the applied thermal gradient $G$ and thus the engulfment velocity $V\propto G$ we can control the relative strength of these two effects and thus the final shape of the droplet.

Since typical time-scales associated with the droplet engulfment in our experiments are extremely slow, much larger than the viscous time-scale $t_v \sim \mu R/\sigma$, the overall process can be treated as quasi-steady. 
Therefore, the unsteady term $\partial_t h$ in Eq.~(\ref{eq:thinfilmeqs}) can be omitted $\left( \partial_t h = 0 \right)$ to determine the steady state of the film thickness as:
\begin{equation}
\begin{split}
h(x) = \frac{h_0}{R} + \left(\frac{L}{R} \right)^3 \frac{Ca}{6 Pr}\zeta(x) \left[\frac{3MaR}{H_0} - \frac{1}{Br} \right] \\ +\frac{1}{2}\frac{p_{\infty}R}{\sigma}\left[\left( \frac{L}{R} \right)^2 - x^2 \right],
\end{split}
\end{equation}
with $\zeta(x) = \left[\left(\frac{xR}{L} \right)^3 - 3\left(\frac{xR}{L} \right)^2 +2 \right]$. 
For this result, we have used the boundary conditions $\partial^2_x h\vert_{{}_{x=L/R}} = -p_{\infty}R/\sigma$, $\partial_x h\vert_{{}_{x=0}}=0$ and $h \vert_{{}_{x=L/R}}= H_0/R$, where $H_0$ is a typical film thickness, $L$ is the length of a stretched droplet, and $p_{\infty}$ a reference ambient pressure.
Notably, this result corroborates with our experimental observations that, promoted by Marangoni flow, thicker liquid films envelop the droplet at higher engulfment velocities.
As the thermal gradient $G$ increases (and thus $V$) the flow towards the colder side is enhanced, leading to the thickening of the liquid film and hence altering the pressure profile within, causing the drop to deform less, whereas intermolecular interactions are responsible for its thinning at lower velocities.
Further, using the relationship between pressure and film thickness, the lubrication pressure within the thin film is estimated to be
\begin{equation}
\Delta p_{\mathrm{lub}} = \frac{\sigma}{R^2}\frac{Ca}{Pr} \left[\frac{3 Ma R}{H_0} - \frac{1}{Br} \right]\left(L - X \right) + \frac{\rho_s \mathcal{L}G}{T_m} X,
\label{eq:presslub}
\end{equation}
where $ \Delta p_{\mathrm{lub}} = p_{\mathrm{film}} - p_{\infty}$.
It is apparent from the above expression that the lubrication pressure within the intervening thin liquid film, irrespective of the velocity of droplet engulfment, varies linearly along the droplet interface.
It also suggests that dynamic Marangoni effects are dominant over intermolecular interactions if $3 Ma\,Br\,R /H_0 < 1$.

To relate the linearly varying pressure in the enveloping thin films with the deformed shape of an engulfed droplet, we use the analogy of a pendant droplet, whose shape is strikingly similar to that of an engulfed droplet.  
As described in Ref.\,\cite{roman2020pendulums}, the canonical interfacial shape of a pendant droplet can be expressed in a non-dimensional form as:       
\begin{equation}
\frac{\mathrm{d}^2\theta}{\mathrm{d}s^2} = - \sin \theta,
\label{eq:pendulumeqs}
\end{equation}      
where $\theta$ is the angle measured through the liquid from the vertical to the tangent, $s$ is the curve-linear coordinate along the interface, and $\mathrm{d}\theta/\mathrm{d}s$ is the local curvature of the interface.
Eq.\,(\ref{eq:pendulumeqs}) simply represents the balance between surface tension and hydrostatic pressure along the interface of a pendant droplet.
Interestingly, the same functional form is also recovered for a droplet enveloped by a thin-film in which the pressure varies linearly.
We refer to Supplemental Material for a detailed derivation.
Accordingly, solving Eq.\,(\ref{eq:pendulumeqs}) for given total energy $\mathcal{E} = 1/2\,\left( \mathrm{d}\theta / \mathrm{d} s \right)^2 - \cos \theta$ yields the deformed shape of the engulfed droplet.
This is equivalent to solving a pendulum equation with total energy $\mathcal{E}$, which turns out to be the only parameter of our model.
Note that the values of the energy parameter $\mathcal{E}$ are chosen so as to exactly match the aspect ratio of the deformed engulfed droplets.
Fig.\,\ref{fig:5} shows an excellent match between the experimentally measured deformed droplet shapes and our model predictions for $\mathcal{E} = 0.615\mathrm{,}\, 0.83$, and $0.99$, respectively.

In summary, in this Letter, we reported that a soft particle when engulfed into a growing crystal sustains mechanical deformations.
These deformations are found to strongly depend on the engulfment velocity and to be mediated by pressure variations in the unfrozen thin liquid films intercalating between the particle and the encapsulating crystal.
In case of droplets or bubbles, the presence of a free-interface evokes thermal Marangoni forces, which critically change the local dynamics.
We demonstrated that the strength of the flow driven by Marangoni stresses dictates the thickness of the intervening liquid film, which in turn controls the deformation of the engulfed particle.\\
Our work quantitatively demonstrates why faster freezing (i.e., larger engulfment velocity $V$) has a much weaker impact on the soft particle's shape than slow-freezing, and what parameters control this process.
These findings are extremely relevant for exerting better control over cyro-preservation procedures of food emulsions and bio-specimen, where mechanical deformations of the dispersed medium are important to the freeze-thaw stability of the solidified material. 

\section*{Acknowledgements}
Authors thank Gert-Wim Bruggert for the technical support in building the experimental setup, and M. S. Saleem, T. T. K. Chan and G. Lajoinie for their help in generating the droplet suspensions. The authors acknowledge the funding by Max Planck Center Twente, NWO, the ERC Adv. Grant No. DDD 740479, and the Balzan Foundation.

\bibliography{references}

\end{document}